\documentclass[aip,graphicx,reprint,longbibliography,noeprint,citeautoscript]{revtex4-1}

\usepackage[pdftex]{graphics}
\usepackage{amssymb,amsmath}
\usepackage{bm}
\usepackage{siunitx}

\newcommand{\kBT}{k_\mathrm{B}T}
\newcommand{\Tr}{\mathrm{Tr}}

\makeatletter
\let\@fnsymbol\@fnsymbol@latex
\@booleanfalse\altaffilletter@sw
\makeatother

\begin{document}

\title{Nuclear quantum effects slow down the energy transfer in biological light-harvesting complexes}

\author{Johan E. Runeson}
\email{johan.runeson@physik.uni-freiburg.de. \\ Present address: Institute of Physics, University of Freiburg, 79104 Freiburg, Germany.}
\affiliation{Physical and Theoretical Chemistry Laboratory, Department of Chemistry, University of Oxford, OX1 3QZ Oxford, United Kingdom}
\author{David E. Manolopoulos}
\affiliation{Physical and Theoretical Chemistry Laboratory, Department of Chemistry, University of Oxford, OX1 3QZ Oxford, United Kingdom}

\begin{abstract}
We assess how quantum-mechanical effects associated with high-frequency chromophore vibrations influence excitation energy transfer in biological light-harvesting complexes. We begin with a mixed quantum-classical theory that combines a quantum description of the electronic motion with a classical description of the nuclear motion in a way that is consistent with the quantum-classical equilibrium distribution. We then include nuclear quantum effects in this theory with a variational polaron transformation of the high frequency vibrational modes. This approach is validated by comparison with fully quantum mechanical benchmark calculations and then applied to three prototypical biological light-harvesting complexes. We find that high-frequency vibrations delay the energy transfer in the quantum treatment, but accelerate it in the classical treatment. For the inter-ring transfer in the light-harvesting complex 2 of purple bacteria, the transfer rate is a factor of 1.5 times slower in the quantum treatment than the classical. The transfer timescale in the Fenna--Matthews--Olson complex is essentially the same in both cases, whereas the transfer in light-harvesting complex II of spinach is 1.7 times slower in the quantum treatment. In all cases, the quantum mechanical long-time equilibrium populations of the chromophores are well reproduced by the classical treatment, suggesting that nuclear quantum effects are generally unimportant for the directionality of energy transfer. Nuclear quantum effects do however reduce the transfer rate in systems with large excitonic energy gaps and strong vibronic coupling to high-frequency vibrational modes.
\end{abstract}

\maketitle

\section*{Introduction}
Photosynthesis is a process of fundamental importance to life on earth and a source of inspiration for the development of solar fuels. However, there are still some aspects of its mechanism that are open to debate,  
including the long-standing question of whether biological light harvesting exhibits any quantum effects.\cite{lambert2013QBreview,higgins2021vibronic,Kim2021quantumbiology} For almost two decades, this debate focused on the interpretation of oscillatory signals observed in nonlinear electronic spectroscopy experiments.\cite{engel2007evidence} These signals were originally attributed to inter-exciton coherences, but a concensus has now emerged that they have a more straightforward vibrational or vibronic origin.\cite{cao2020review,Schultz2024review} In recent years, the debate has therefore shifted to consider the quantum effects associated with the vibrations themselves,\cite{Kundu2022science} which we shall refer to here as ``nuclear quantum effects". Our goal in this article is to contribute to the debate by comparing classical and quantum treatments of the vibrational modes in explicit simulations of the energy transfer in biological light-harvesting complexes. 

On the one hand, a series of recent articles have argued that nuclear quantum effects are essential for the function of photosynthetic antenna complexes. For a vibronic model of the light harvesting complex 2 (LH2) in purple bacteria, Kundu and Makri\cite{Kundu2022science} found that a classical treatment of vibrations provided a poor description of exciton equilibration compared to a fully quantum treatment, which led them to conclude that quantum nuclear motion is necessary for the correct flow of energy. Similar statements have been made by other authors,\cite{mancal2013,brumer2018vibrational,renger2021detailedbalance} and two recent reviews\cite{cao2020review,Schultz2024review} have identified nuclear quantum effects as responsible for the existence of an energy funnel and therefore for the directionality of energy transfer. On the other hand, numerous simulations with classical nuclei have been shown to be consistent with the exact quantum dynamics of models of the Fenna--Matthews--Olson (FMO) complex of green sulfur bacteria,\cite{tao2010FMO,coker2016photosynthesis,cotton2016lightharvest,saller2020faraday,runeson2022fmo,Lawrence2024sizeconsistent} suggesting that nuclear quantum effects do not play any role in its energy transfer. Hence the literature is still divided on the issue.

To explain why this is, we first need to clarify what we mean by a nuclear quantum effect in the present context. Light-harvesting complexes enable efficient excitation energy transfer through an interplay of dipolar coupling between chromophores and vibronic coupling to nuclear motion. In a fully quantum description, the nuclear degrees of freedom are quantized in the same way as the excitonic degrees of freedom. However, fully quantum dynamics is computationally demanding and only possible for simple model systems, which has motivated the development of approaches that treat the nuclear degrees of freedom classically. An important issue here is that the classical treatment is not unique. Some classical formulations are better justified than others, and different approaches can be more or less accurate in comparison to the quantum description. So only when the best available classical treatment is unable to reproduce the quantum result does it make sense to identify a quantum effect. Such an identification indicates the level of theory that is required -- or sufficient -- to accurately describe the phenomenon. If one can reproduce the quantum result with a classical simulation then there is no quantum effect. 


The standard quantum description of exciton energy transfer is based on the model Hamiltonian $H = H_\mathrm{s} + H_\mathrm{b} + H_\mathrm{sb}$, where $H_\mathrm{s}$ is the Hamiltonian of the excitonic system, $H_\mathrm{b}$ that of a vibrational bath, and $H_\mathrm{sb}$ that of the system-bath interaction. In a basis of locally excited chromophores (or `sites'), the first term is written as
\begin{equation}
    H_\mathrm{s} = \sum_n \varepsilon_n |n\rangle \langle n| + \sum_{n\neq m} J_{nm} |n\rangle \langle m|,\label{eq:Hex}
\end{equation}
where $\varepsilon_n$ are the site energies and $J_{nm}$ the inter-site couplings. The other terms,
\begin{equation}
    H_\mathrm{b} = \sum_{nk}\hbar\omega_k \left(b_{nk}^\dagger b_{nk}+\tfrac{1}{2}\right)
\end{equation}
\begin{equation}
    H_\mathrm{sb} = \sum_{nk}\hbar\omega_k g_{k}(b_{nk}^\dagger+b_{nk})|n\rangle\langle n|,
\end{equation}
model how the site energies fluctuate due to interaction with vibrational degrees of freedom with frequencies $\omega_k$ and coupling strengths $g_k$. As is customary, we assume that the sites are coupled to identical and independent baths. 
Typically, one can attribute the high-frequency bath modes to intramolecular vibrations and the low-frequency modes to phonons of the solvent and protein scaffolding. 

To identify the role of nuclear quantum effects in this model, one first needs to define an appropriate classical limit for the nuclear motion. 
Based on past literature, it appears that there are many mixed quantum-classical approaches to choose from: surface hopping,\cite{tully1990hopping} Ehrenfest dynamics, phase-space mappings,\cite{stock2005nonadiabatic,miller2016review} quantum-classical path integrals,\cite{LambertMakri2012} etc.
Each of these methods has its own definition of the force acting on the classical nuclei. The excitonic state-dependent part of this force is the so-called `back-action' of the quantum system on the classical degrees of freedom. As has been discussed elsewhere\cite{runeson2022fmo}, part of the debate about nuclear quantum effects is semantic in the sense that some authors use the term `classical' to mean `classical with no back-action', or  `classical with disregard for Newton's third law'. It is well known that no back-action leads to equal long-time populations of all quantum states, an unphysical situation that corresponds to elevating the excitonic system to infinite temperature. Neglecting the back-action is therefore inconsistent with the quantum-classical equilibrium distribution at the temperature of interest.\cite{runeson2022fmo} 

The quantum mechanical Boltzmann population of site $n$ is
\begin{equation}\label{eq:Qeq}
    \langle |n\rangle\langle n|\rangle = \frac{\Tr_\mathrm{nuc}[\Tr_\mathrm{ex}[e^{-\beta H}|n\rangle\langle n|]]}{\Tr_\mathrm{nuc} [\Tr_\mathrm{ex}[e^{-\beta H}]]},
\end{equation}
where the labels `nuc' and `ex' refer to the nuclear and excitonic degrees of freedom, respectively.
The classical analog of the nuclear trace is obtained by replacing $b_{nk}=\frac{1}{\sqrt{2\hbar\omega_k}}(\omega_k q_{nk}+ip_{nk})$ and treating $(p,q)=(\{p_{nk}\},\{q_{nk}\})$ as classical variables. The classical limit of the population then emerges from an average of the excitonic Boltzmann distribution over the nuclear phase space,
\begin{equation}\label{eq:mixedQC}
    \langle |n\rangle\langle n|\rangle_{\mathrm{cl}} = \frac{\int dp dq \,\Tr_\mathrm{ex}[e^{-\beta H(p,q)}|n\rangle\langle n|]}{\int dp dq \,\Tr_\mathrm{ex}[e^{-\beta H(p,q)}]}.
\end{equation}

Even though the mixed quantum-classical statistics in Eq.~\eqref{eq:mixedQC} is much simpler than the fully quantum statistics in Eq.~\eqref{eq:Qeq}, very few mixed quantum-classical methods are actually consistent with it. Ehrenfest dynamics is known to overheat the excitonic subsystem in much the same way as neglecting back-action. Fewest-switches surface hopping often gives more accurate long-time populations than Ehrenfest dynamics, but the stochastic nature of this algorithm makes it hard to analyse the circumstances under which they can be trusted. A recently developed alternative is Mannouch and Richardson's deterministic `mapping approach to surface hopping' (MASH).\cite{Mannouch2023mash} We have adapted this method to handle multiple electronic states,\cite{Runeson2023mash} successfully applied the adaptation to the energy transfer in FMO,\cite{Runeson2023pccp} and proven that it exactly recovers the correct quantum-classical equilibrium populations in Eq.~\eqref{eq:mixedQC}.\cite{Runeson2023mash} Since our multi-state version of MASH is the only method we are aware of that both recovers these equilibrium populations and has been shown to work well for excitonic Hamiltonians, it is the method we shall use for the present calculations.

\section*{Results}
Of the many known biological light-harvesting complexes,
we have selected three representative examples: LH2 in purple bacteria, FMO in green sulfur bacteria, and the light-harvesting complex II (LHCII) in spinach. All three systems have been studied extensively, both experimentally and theoretically, and the historical  debate about nuclear quantum effects has largely been focused on LH2 and FMO. We will consider the physiologically relevant temperature of $\SI{300}{K}$ in all of our simulations.

\subsection*{Light-harvesting complex 2 in purple bacteria}
\begin{figure}
    \centering
    \includegraphics{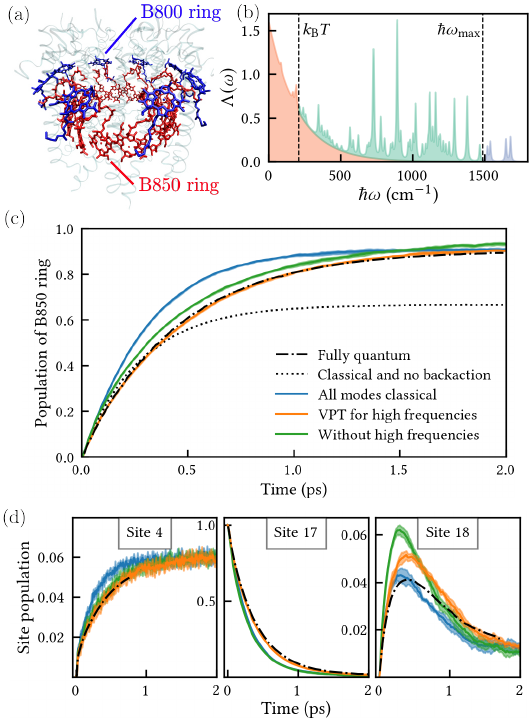}
    \caption{Exciton energy transfer in LH2. (a) The LH2 structure, composed of the B850 ring (red, sites 1-16) and the B800 ring (blue, sites 17-24) embedded in a protein scaffolding (shaded). (b) Density of bath reorganization energies, $\Lambda(\omega)=\sum_k \hbar\omega_k g_k^2 \delta(\hbar\omega-\hbar\omega_k)$. This dimensionless quantity is partitioned into a low-frequency part that contains the solvent and discrete modes below $\kBT$ (red), a high-frequency part containing discrete modes between $\kBT$ and $\hbar\omega_\mathrm{max}$ (teal), and an off-resonant part containing disrete modes above $\hbar\omega_\mathrm{max}$ (purple). For visual purposes, each discrete mode has been broadened by $\gamma=\SI{10}{cm^{-1}}$ such that its contribution to $\Lambda(\omega)$ is $\frac{2}{\pi}g_k^2 \omega_k^2 \frac{\gamma \omega}{(\omega_k^2-\omega^2)^2+\omega^2\gamma^2}$.
    (c) Time-dependent population of the B850 ring after an initial excitation on site 17 in the B800 ring. Shades around lines indicate two standard errors in the mean. (d) Time-dependent populations of a few representative sites.}
    \label{fig:lh2}
\end{figure}

As our first example, we consider the light-harvesting complex 2 (LH2) in purple bacteria. This complex consists of two bacteriochlorophyll rings, which have absorption maxima near 800 nm and 850 nm and are consequently called the B800 and B850 rings. Its structure varies across different species\cite{Schlau-Cohen2020experiment}, and in the following we focus on the octameric complex found in \emph{Magnetospirillum molischianum} (note that the former generic names \emph{Rhodospirillum} and \emph{Phaeospirillum} are often still used in the literature). In total, the complex comprises 24 chromophores embedded in a protein scaffolding, as depicted in Fig.~\ref{fig:lh2}(a) (using the crystal structure from Ref.~\onlinecite{Koepke1996lh2}). The primary function of the complex is to funnel excitation energy inwards, from B800 to B850, in order to reach a reaction centre near the centre of the B850 ring.

As a fully quantum benchmark for the exciton dynamics we use the recent path-integral calculations by Kundu and Makri.\cite{Kundu2022science} Their model Hamiltonian uses site energies and couplings from Tretiak \emph{et al}.,\cite{Tretiak2000LH2_second} and a bath spectral density that combines an experimentally determined intramolecular contribution\cite{Raetsep2011Bchla} with a phenomenological solvent contribution (see Supplementary Note 1 for full details).
To illustrate the spectral density, Figure~\ref{fig:lh2}(b) shows the (dimensionless) density of bath reorganization energies, $\Lambda(\omega) = \sum_k \hbar\omega_k g_k^2 \delta(\hbar\omega-\hbar\omega_k)$. [Note that other common ways to represent the spectral density are through the density of Huang-Rhys factors, $S(\omega)=\sum_k g_k^2\delta(\omega-\omega_k)$, and through the spectral function $J(\omega)=\pi \sum_k \hbar\omega_k^2g_k^2\delta(\omega-\omega_k)$, both of which have the same information content as $\Lambda(\omega)$. We prefer to use $\Lambda(\omega)$ because it is more transparently related to the total bath reorganisation energy, $\Lambda = \hbar\int_0^{\infty} \Lambda(\omega)\,d\omega$.]

After an initial excitation on the B800 ring (assumed to be localized on site 17), the transfer to the B850 ring can be described by rate-like kinetics, as shown in Fig~\ref{fig:lh2}(c). In the fully quantum calculation\cite{Kundu2022science} (dash-dotted line), the B850 population reaches a long-time value of $P_\infty=0.90$ with time constant $\tau=\SI{0.45}{ps}$, as determined from a fit to $P_{\rm B850}(t)=P_\infty(1-e^{-t/\tau})$. Kundu and Makri also simulated the dynamics in the limit of classical nuclei, with a method that excludes back-action (shown as the dotted line in Fig.~1(c)).\cite{Makri2015qcpi,Kundu2022science} As we have already mentioned, such a classical approach leads to equal long-time populations on all sites, reducing the long-time yield of the B850 ring to $16/24=2/3$. 
Since this is less than the quantum yield of 90\%, Kundu and Makri concluded that quantum nuclear motion was responsible for the more efficient energy transfer seen in their path integral calculation.\cite{Kundu2022science,Schultz2024review}

We have included back-action in the classical calculation by running mixed quantum-classical dynamics with MASH. The result of this calculation is shown as the solid blue line in Fig.~1(c). MASH is seen to give an accurate long-time B850 yield despite treating the nuclear motion classically. This implies that there is little difference between the fully quantum [Eq.~\eqref{eq:Qeq}] and mixed quantum-classical [Eq.~\eqref{eq:mixedQC}] equilibrium distributions for this problem, as can be verified with a simple equilibrium simulation (see Supplementary Note 2). It also shows that it is unreliable to draw conclusions about nuclear quantum effects from comparisons with classical simulations without back-action.\cite{Kundu2022science,cao2020review} When back-action is included in the classical simulation, there is no longer any evidence for quantum nuclear motion leading to more efficient energy transfer in the sense of giving a larger B850 yield.

The present calculations do, however, reveal a different nuclear quantum effect. The  time constant of the fully quantum calculation in Fig. 1(c) ($\tau=0.45$ ps) is almost 50\,\% larger than that of the ``all modes classical" MASH calculation ($\tau=0.31$ ps). Presumably this is because a significant fraction of the reorganisation energy in LH2 comes from vibrational modes with $\hbar\omega> k_{\rm B}T$ [see Fig. 1(b)], the classical treatment of which neglects their zero point energy. What is less obvious is why switching to a quantum treatment {\em decreases} the energy transfer rate. A clue is provided by the similarity between the present excitonic model and the Holstein models that are widely used to describe charge transport in organic semiconductors. In that context, the primary effect of high-frequency on-site modes is to narrow the band width and thereby reduce the charge mobility,\cite{Nematiaram2020perspective,Fetherolf2020prx,Wang2020vpt} so perhaps a similar effect is in operation here.

To test this hypothesis, we have used a variational polaron transformation\cite{Yarkony1976,Wang2020vpt} (VPT, see Methods) to obtain a set of dressed excitonic states that mix bare excitonic states with nuclear modes (the selection of modes will be made precise below). The transformed Hamiltonian is also of system-bath form, $H' = H'_\mathrm{s} + H'_\mathrm{b} + H'_\mathrm{sb}$, where $H'_\mathrm{s}$ is a new system Hamiltonian with renormalized site energies ($\varepsilon_n'$) and couplings ($J_{nm}'$), $H'_\mathrm{b}=H_\mathrm{b}$ is unchanged, and $H'_\mathrm{sb}$ contains reduced linear couplings on the diagonal and additional exponential couplings on the off-diagonal. The transformation is variationally optimized to minimize the contribution of $H'_\mathrm{sb}$ to the free energy.\cite{Yarkony1976}
To account for the remaining bath coupling, other authors have suggested using a perturbative master equation with $H'_\mathrm{sb}$ as the perturbation.\cite{Jang2008polaron,McCutcheon2011,Zimanyi2012,Kimura2014variational,Lee2015polaron} However, such an approach is not ideal for the present system for the following reasons. The fluctuations in
$H'_\mathrm{sb}$ are only perturbatively small when the temperature is low (frequencies are high) and the Huang-Rhys factors are sufficiently small. For high temperature (low frequencies) or large Huang-Rhys factors,
the renormalized couplings $J'_{nm}$ approach zero and the corresponding fluctuations in $H'_\mathrm{sb}$ are large.\cite{Troisi2010polaron} 
Low-frequency modes are therefore already better described with MASH (which is non-perturbative and non-Markovian\cite{Runeson2023pccp}) so they should not be included in the VPT. The renormalized system Hamiltonian captures the main effect of the high-frequency modes, and in accordance with common practice for organic semiconductors, we neglect the effect of their remaining fluctuations (which is small compared to the effect of the low-frequency modes). 

A natural energy scale to distinguish between `low' and `high' frequencies is $\kBT$. However, it is not ideal to introduce a sharp cut-off in the spectral density, since this corresponds to a bath with unphysically long-lived correlations.\cite{Berkelbach2012hybrid} A pragmatic solution is to make use of the existing division between the smooth solvent part and the discrete intramolecular part of the spectral density. In the following, we 
define the `low-frequency part' to consist of (i) discrete modes with $\hbar\omega_k<\kBT$ and (ii) the entire solvent part of the spectral density, assuming it is smooth and dominated by frequencies lower than $\kBT$. This is the part of the spectral density that is coloured red in Fig.~\ref{fig:lh2}(b). The `high frequency' part that is included in the VPT is the remainder, which only contains discrete modes with frequencies $\hbar\omega_k>k_{\rm B}T$. This can be further divided into two contributions: the region couloured in teal between $k_{\rm B}T$ and $\hbar\omega_{\rm max}$, and the region couloured in purple beyond $\hbar\omega_{\rm max}$, where $\hbar\omega_{\rm max}$ is the maximum gap between the eigenenergies of $H_{\mathrm{s}}$ ($\SI{1488}{cm^{-1}}$ for LH2). The reason for this further subdivision is that the purple part of the spectral density is out of resonance with the exciton dynamics, and so is unlikely to be responsible for any interesting effects. Indeed, classical MASH gives the same results when it is run both with and without these non-resonant modes.


The MASH dynamics after the VPT has been applied (orange line in Fig.~\ref{fig:lh2}(c)) is in close agreement with the fully quantum result. This is also the case for the site-specific populations shown in panel (d), with the exception of site 18, which has a turnover where our method overestimates the maximum population. Overall, the VPT has corrected the timescale of the dynamics compared to treating all modes classically. 
We obtain band-narrowing factors $J_{nm}'/J_{nm}$ that are on average 0.89 (arithmetic average over sites) and minimally 0.80, corresponding to a mild reduction in the couplings.
These band-narrowing factors account for quantum nuclear statistics by virtue of the factor of $\coth(\beta\hbar\omega_k/2)=2n_k+1$ in Eq.~\eqref{eq:Bnm}, where $n_k=(e^{\beta\hbar\omega_k}-1)^{-1}$ is the Bose--Einstein distribution (see Methods). 
Since the only difference between the blue and orange lines in Fig.~1 is whether the high-frequency modes are treated classically (with MASH) or quantum-mechanically (with the VPT), we can identify the change in rate as a nuclear quantum effect.
In the classical simulation, the high-frequency modes are actively promoting the energy transfer, whereas in the quantum calculation they are largely frozen in their ground states. Indeed, compared to a MASH calculation without the high-frequency modes (the green lines in Fig.~1), treating these modes classically enhances the rate of energy transfer, whereas treating them quantum mechanically inhibits it.

Before continuing to the next system, we should point out that a more popular way to include nuclear quantum effects in system-bath problems is to start the bath variables from a Wigner distribution. When we do this, we find that the resulting MASH dynamics (the gray line in Supplementary Figure~\ref{fig:wigner}) differs from the classical MASH
dynamics and is consistent at short times with the slower energy transfer obtained using the VPT. However, since the Wigner distribution effectively initializes the high-frequency modes at an elevated temperature, the resulting long-time populations correspond to an overheated equilibrium. 
For this reason, using an initial Wigner distribution in MASH is less accurate at long times than combining it with the VPT.


\subsection*{Fenna--Matthews--Olson complex}
Our identification of nuclear quantum effects in LH2 stands in contrast to previous studies of another well-studied light-harvesting system, the Fenna--Matthews--Olson (FMO) complex in green sulfur bacteria. For simple FMO models in which the bath is treated as a single overdamped oscillator, many classical trajectory methods (including MASH\cite{Runeson2023mash} and others\cite{coker2016photosynthesis,cotton2016lightharvest,runeson2020}) have been found to give results in good agreement with fully quantum dynamics\cite{ishizaki2009FMO}. The obvious question, therefore, is why are nuclear quantum effects more significant in LH2 than in FMO?

\begin{figure}
    \centering
    \includegraphics{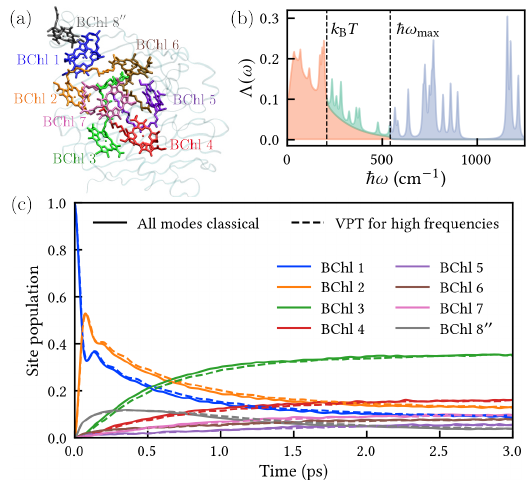}
    \caption{Energy transfer in FMO. (a) Site labelling. (b) Spectral density plotted as in Fig.~\ref{fig:lh2}. (c) The MASH population dynamics is essentially the same with or without the variational polaron transformation.}
    \label{fig:fmo}
\end{figure}

To answer this question, we consider a standard 8-site model of FMO in \emph{Prosthecochloris aestuarii} with site energies and couplings taken from Schmidt am Busch \emph{et al.}\cite{amBusch2011fmo} The site labelling of the complex is shown in Fig.~\ref{fig:fmo}(a). The spectral density consists of a continuous intermolecular part\cite{renger2002,adolphs2006fmo} and a discrete intramolecular part obtained from flourescence line-narrowing experiments.\cite{wendling2000electron} Using the same criteria as for LH2, Fig.~\ref{fig:fmo}(b) shows the division of the spectral density into a low-frequency part (red), a high-frequency part (teal), and a non-resonant part with frequencies larger than $\hbar\omega_\text{max}=\SI{539}{cm^{-1}}$ (purple). Important observations are that (i) the maximum excitonic gap $\hbar\omega_\text{max}$ is smaller for FMO than for LH2, primarily because the inter-site couplings are smaller, and (ii) the vibronic coupling is weaker in FMO than in LH2 in the region between $k_{\rm B}T$ and $\hbar\omega_{\rm max}$. As a result, all the band-narrowing factors $J_{nm}'/J_{nm}$ are $\geq 0.94$, and the dynamics hardly changes when the high-frequency modes are treated  with the VPT instead of classically.
To check that the results are insensitive to the details of the spectral density, we have repeated the calculation with a more detailed intermolecular spectrum extracted from a normal-mode analysis of the full atomistic complex,\cite{klinger2020fmo} and reached the same conclusion (see Fig.~\ref{fig:fmoextra}). Based on these results, we conclude that nuclear quantum effects are negligible in FMO because of its small excitonic energy gaps and its weak vibronic coupling in the relevant `quantum' region between $k_{\rm B}T$ and $\hbar\omega_{\rm max}$.

\subsection*{Light-harvesting complex II in spinach}

\begin{figure}
    \centering
    \includegraphics{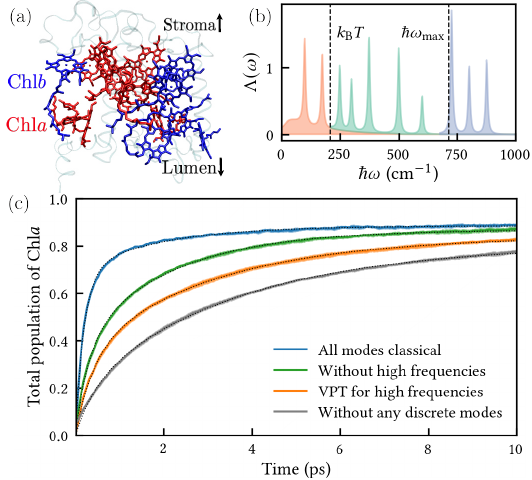}
    \caption{Exciton dynamics in the major LHCII complex of spinach. (a) Structure\cite{Liu2004nature} showing the assignments of Chl$a$ and Chl$b$ chlorophylls in red and blue, respectively, and the orientation relative to the stromal and lumenal layers. (b) Spectral density plotted as in Fig.~\ref{fig:lh2}. (c) Time-dependent population of the Chl$a$ sites (dotted lines are biexponential fits).}
    \label{fig:lhcii}
\end{figure}

Having found that nuclear quantum effects are noticeable in LH2 but negligible in FMO, we now ask which of these two pictures is likely to be more representative of photosynthesis in nature? To answer this question, we have chosen to consider the LHCII complex, which is present in more than 50\,\% of all plants,\cite{blankenship2002} as our third example.
Fig.~\ref{fig:lhcii}(a) shows the major LHCII complex in spinach (\emph{Spinacia oleracea}).\cite{Liu2004nature} It contains 14 chlorophyll sites, which are traditionally divided into two groups, Chl$a$ and Chl$b$. The latter have higher excitation energies, resulting in an energy funnel that is directed towards the Chl$a$ sites.

Many different models of this complex have been proposed in the literature. Here, we use the site couplings of M\"uh \emph{et al.},\cite{Mueh2010lhcii} with refined site energies from Ref.~\onlinecite{Mueh2012refined} and a spectral density from Ref.~\onlinecite{Renger2011lhc2} [shown in Fig.~\ref{fig:lhcii}(b)]. This model reproduces the experimental linear absorption, linear dichroism, and flourescence spectra. 
Although there are alternative models that would provide quantum benchmarks,\cite{kreisbeck2014LH2,Leng2018lhcii,Novoderezhkin2023} we have chosen not to compare with them because their spectral density consists of a single overdamped oscillator dominated by frequencies larger than $\kBT$, which is inconsistent both with our criteria for using the VPT and with the experimental low-temperature flourescence spectrum.\cite{Mueh2010lhcii}
For the present model, $\hbar\omega_\text{max}=\SI{713}{cm^{-1}}$ is intermediate between the case of LH2 ($\SI{1488}{cm^{-1}}$) and FMO ($\SI{539}{cm^{-1}}$). The same holds for the fraction of reorganization energy in the interval $[\kBT,\hbar\omega_{\rm max}]$ relative to that in $[0,\hbar\omega_{\rm max}]$,
which is 44\,\% for LHCII, 11\,\% for FMO, and 54\,\% for LH2. 
For the purpose of qualitatively assessing nuclear quantum effects, it is therefore justified to use the same analysis as in the previous examples without reference to another fully quantum benchmark.

For simplicity, we start the simulation from the highest energy site, which is labelled $b$609 in the protein database entry 1RWT.\cite{Liu2004nature} For each of our methods, the resulting downhill energy transfer to the Chl$a$ sites can be fitted to a sum of two exponentials with time constants $\tau_1$ and $\tau_2$ (the fits are shown with dotted lines in Fig.~\ref{fig:lhcii}(c)). When all modes are treated classically with MASH, we obtain $\tau_1=\SI{0.20}{ps}$ and $\tau_2=\SI{1.8}{ps}$. When the high-frequency modes are handled with the VPT, the corresponding results are $\tau_1=\SI{0.40}{ps}$ and $\tau_2=\SI{3.0}{ps}$. Expressed in terms of rates, the associated `classical' rate constants are 2.0 and 1.7 times larger, respectively, than those obtained with the VPT. In other words, the time delay associated with nuclear quantum effects is even more pronounced for LHCII than for LH2. As was the case for LH2, a comparison to a calculation without the high-frequency modes (green line) shows that the classical treatment speeds up the transfer whereas the quantum treatment (VPT) slows it down. This last observation appears at first glance to contradict a previous study\cite{Renger2011lhc2} which found that the discrete part of the spectral density increases the rate. However, if we neglect all discrete modes (gray line) the rate is significantly slower again, so our results do not contradict this previous study, but rather indicate that the speedup is caused mainly by low-frequency discrete modes.

\section*{Discussion}
In conclusion, the main difference between quantum and classical treatments of vibrational motion in light-harvesting complexes is \emph{not} the direction of energy flow, but instead the rate of energy transfer.
More precisely, it is sufficient to treat the vibrational motion classically to obtain the correct long-time equilibrium, but the rate of transfer will generally be overestimated.
Renormalizing the Hamiltonian with the VPT provides a simple way to account for this quantum effect.
For biologically relevant parameter regimes, the quantum effect on the rate is benign (less than a factor 2), and the delay may be regarded as a quantum correction to an otherwise qualitatively valid classical picture. 

In reaching these conclusions, we have assumed that the coupling to high-frequency vibrations is weak enough to discard the fluctuations of the renormalized Hamiltonian. This assumption is more questionable for processes involving charge transfer (e.g. in the reaction centre), but for such processes the intersite coupling is typically small enough to use Förster theory (the perturbative master equation that results from a full polaron transformation). In comparison to classical Marcus theory, Förster theory can account for both positive and negative quantum corrections to the transfer rates, and the effect can reach orders of magnitude in the inverted regime. 

For the present systems, the quantum correction to the energy transfer rate is so slight that the majority of it can be captured simply by neglecting the high-frequency modes rather than treating them classically. In organic semiconductors, the slowdown of the charge mobility due to high-frequency ``killer modes'' can be much more dramatic.\cite{Dettmann2023killer} It is therefore conceivable that chlorophylls have been selected during the evolution of light-harvesting complexes because their excitons are relatively weakly coupled to high-frequency modes.

One of the motivations for studying biological energy transfer is as inspiration for the design of synthetic complexes such as porphyrin nanostructures.\cite{Yong2015synthetic,Pannwitz2019,anderson2020nanoring,Keijer2021review} 
However, since the porphyrins in these nanostructures are connected by covalent bonds, they may be more strongly affected by high-frequency vibrations than the chlorophylls we have considered here. Our results for biological complexes show that care needs to be taken when treating high-frequency vibrations, and that the VPT provides a simple way to capture their quantum mechanical behaviour. We can see no reason why this would not also be the case for synthetic complexes.

\section*{Methods}
\subsection*{Multi-state mapping approach to surface hopping}
In this mixed quantum-classical method, one starts by replacing the nuclear operators by their classical limit, $b_{nk}=\frac{1}{\sqrt{2\hbar\omega_k}}(\omega_k q_{nk}+i p_{nk})$, leading to a multi-state Hamiltonian of the (mass-scaled) form 
\begin{equation}
    H(p,q) = \frac{p^2}{2} + V(q),
\end{equation}
\begin{equation}
    V(q) =  \sum_{n,m} V_{nm}(q)|n\rangle\langle m|.
\end{equation}
The goal is to propagate trajectories of the nuclear variables $(p,q)$ alongside the wavefunction $|\psi\rangle=\sum_n c_n|n\rangle$, where one may think of the complex coefficients $c_n$ as phase-space variables for the excitonic degrees of freedom. Multi-state MASH does this by using the force of the local eigenstate with the largest instantaneous population to propagate the nuclei.\cite{Runeson2023mash} At each timestep, the local eigenstates are found by solving the eigenvalue problem
\begin{equation}
    V(q)|a(q)\rangle = V_a(q)|a(q)\rangle
\end{equation}
and their populations are computed as $P_a = |\langle \psi|a(q)\rangle|^2$. The force can then be written as
\begin{equation}
    \dot{p} = -\sum_a \langle a(q)|\nabla V(q)|a(q)\rangle \Theta_a(P),
\end{equation}
where 
\begin{equation}
    \Theta_a(P)=\begin{cases}
        1 \text{ if } P_a>P_b~ \forall\, b\neq a \\
        0 \text{ otherwise}.
    \end{cases}
\end{equation}
Only the maximally populated state has a non-zero value of $\Theta_a(P)$, and we refer to this state as the `active' state.
Whenever a new state reaches a higher population than the previously active state, there is an associated jump in the potential energy, and the momentum is adjusted to preserve the total energy. The momentum is rescaled along the particular direction specified in Appendix~E of Ref.~\onlinecite{Runeson2023mash}. If the available kinetic energy is insufficient to hop, the momentum is instead reversed along the same direction (analogous to a particle bouncing off a wall \cite{Mannouch2023mash}).

The measure populations, we use the estimator\cite{Runeson2023mash} 
\begin{equation}
    \Phi_n = \frac{1}{N} + \alpha_N\left(|c_n|^2-\frac{1}{N}\right)
\end{equation}
where $\alpha_N=(N-1)/(H_N-1)$ and $H_N=\sum_{n=1}^N \frac{1}{n}$. This estimator is constructed to be consistent with the mixed quantum-classical equilibrium population in Eq.~\eqref{eq:mixedQC}, which the `Ehrenfest population' $|c_n|^2$ is not. Various initial conditions for the coefficients $c_n$ are compatible with this estimator.\cite{Runeson2023pccp} In the present calculations we used  the `focused' initial condition in which $c_n=\sqrt{r_n}e^{i\phi_n}$, where $\phi_n$ is sampled uniformly from $[0,2\pi)$ and $r_n$ is chosen so that $\Phi_n$ is 1 for the initial state and 0 for all other states. If $i$ is the initial state, one obtains $r_i=\frac{N+\alpha_N-1}{N\alpha_N}$ and $r_{n\neq i}=\frac{\alpha_N-1}{N\alpha_N}$. 
The nuclear variables were initialized from a classical Boltzmann distribution centred at $q=0$, corresponding to a vertical excitation from the ground-state equilibrium.
The results were averaged over $10^5$ trajectories for LH2 and LHCII, and $5\times 10^4$ trajectories for FMO. Each simulation was divided into 5 batches, from which the standard errors in the mean were calculated.

\subsection*{Variational polaron transformation}
The variational polaron transformation (VPT)\cite{Yarkony1976} is an exact canonical transformation $H' = e^G H e^{-G}$ with the generator
\begin{equation}
    G = \sum_{nk}f_{nk}(b_{nk}^\dagger-b_{nk})|n\rangle\langle n|.
\end{equation}
The transformation leads to another system-bath Hamiltonian $H'=H'_\mathrm{s}+H'_\mathrm{b}+H'_\mathrm{sb}$, where 
\begin{equation}
    H'_\mathrm{s}= \sum_n \varepsilon'_n |n\rangle\langle n| + \sum_{n\neq m}J_{nm}'|n\rangle\langle m| 
\end{equation} 
describes an excitonic system with renormalized site energies and couplings, $H'_\mathrm{b}=H_\mathrm{b}$, and
\begin{multline}\label{eq:Hsb_prime}
    H'_\mathrm{sb} = \sum_{nk}\hbar\omega_k(g_{k}-f_{nk})(b_{nk}^\dagger+b_{nk})|n\rangle\langle n| \\ + \sum_{n\neq m} J_{nm}(B_{nm}-\langle B_{nm}\rangle) |n\rangle\langle m|
\end{multline}
is a renormalized system-bath interaction with $B_{nm} = e^{+\sum_{nk}f_{nk}(b_{nk}^\dagger-b_{nk})-\sum_{mk}f_{mk}(b_{mk}^\dagger-b_{mk})}$.
The renormalized site energies and couplings are
\begin{subequations}
\begin{align}
    \varepsilon'_n &= \varepsilon_n - \sum_k \hbar\omega_k(2f_{nk}g_k - f_{nk}^2) \\
    J_{nm}' &= J_{nm} \langle B_{nm}\rangle \label{eq:Jnmprime}
\end{align}
\end{subequations}
where
\begin{equation}\label{eq:Bnm}
    \langle B_{nm}\rangle = e^{-\frac{1}{2}\sum_k(f_{nk}^2+f_{mk}^2)\coth(\beta\hbar\omega_k/2)}
\end{equation}
is the band narrowing factor.

The aim is to adjust the variational parameters $f_{nk}$ such that $H_1 = H'_\mathrm{sb}$ can be neglected next to $H_0\equiv H'_\mathrm{s}+H'_\mathrm{b}$. 
According to Bogoliubov's theorem, the free energy of the total system is bounded from above by the free energy of the uncoupled system, $F\leq F_0 + \langle H_1\rangle_0$, where $\langle H_1\rangle_0=0$ by construction for any $f_{nk}$. To make $F_0$ as close as possible to $F$, one therefore chooses $f_{nk}$ such that $F_0$ is minimized. In practice, this leads to the condition
\begin{equation}
    0 = \frac{\partial F_0}{\partial f_{nk}} = \sum_M P_M \frac{\partial \varepsilon_M}{\partial f_{nk}}, 
\end{equation}
where $\varepsilon_M=\langle M|H'_\mathrm{s}|M\rangle$ are eigenenergies of $H'_\mathrm{s}$ with eigenstates $|M\rangle$, and $P_M = e^{-\beta \varepsilon_M}/\sum_{N} e^{-\beta \varepsilon_{N}}$. Using the Hellmann--Feynman theorem, $\frac{\partial}{\partial\lambda}\langle M|H|M\rangle=\langle M|\frac{\partial H}{\partial \lambda}|M\rangle$, one obtains
\begin{equation} \label{eq:fnk}
    f_{nk} = \frac{\hbar\omega_k g_k }{\hbar\omega_k   - \coth(\beta\hbar\omega_k/2)\sum_{m}' J_{nm}' \Gamma_{nm} }
\end{equation}
where $\Gamma_{nm}=\sum_M P_M U_{nM}U_{mM}/\sum_M P_M U_{nM}^2$ with $U_{nM}=\langle n|M\rangle$ and the prime on the sum over $m$ indicates that terms with $m=n$ are excluded. This expression differs from  Ref.~\onlinecite{Wang2020vpt} because the exciton states follow a Boltzmann distribution instead of a Fermi--Dirac distribution. Equation~\eqref{eq:fnk} was solved self-consistently for $f_{nk}$ and a stable solution was found within 10 iterations starting from $f_{nk}=g_k$. We verified that the solution was insensitive to the choice of initial guess for all systems considered here.

\section*{Data availability}

\section*{Acknowledgements}
J.E.R. acknowledges funding from the Swiss National Science Foundation (grant no. P500PN\_206641 / 1) and academic support from a Junior Research Fellowship at Wadham College. 

\bibliography{runerefs}

\newpage

\appendix
\onecolumngrid

\renewcommand\thefigure{S\arabic{figure}}  
\setcounter{figure}{0} 
\renewcommand\theequation{S\arabic{equation}}  
\setcounter{equation}{0} 
\setcounter{page}{1}

\section*{Supplementary note 1: Model details}
\subsection*{LH2} The supplementary file \texttt{LH2-HS.dat} contains the $24\times 24$ matrix of site energies and couplings\cite{Tretiak2000LH2_second} (in cm$^{-1}$). The file \texttt{LH2-specden.dat} contains a table with the discrete part of the spectral density\cite{Raetsep2011Bchla} with columns $\hbar\omega_k$ (in cm$^{-1}$) and $g_k^2/1000$ (dimensionless), corresponding to a total reorganization energy of $\SI{217}{cm^{-1}}$. These files were kindly provided to us by Kundu and Makri.\cite{Kundu2022science}
The continuous contribution to the spectral density has the phenomenological form\cite{Kundu2022science} 
\begin{equation} 
J(\omega)=\pi\sum_k\hbar\omega_k^2g_k^2\delta(\omega-\omega_k)=2\pi\xi \hbar\omega e^{-\omega/\omega_\text{c}},
\end{equation}
where $\xi=0.4$ and $\omega_\text{c}=\SI{200}{cm^{-1}}$, which is equivalent to
\begin{equation}
    \Lambda(\omega)=\sum_k \hbar\omega_k g_k^2\delta(\hbar\omega-\hbar\omega_k) = \frac{J(\omega)}{\pi\hbar\omega} = 2\xi e^{-\omega/\omega_\text{c}}
\end{equation}
and contributes a total reorganization energy of $\SI{160}{cm^{-1}}$.
In practice, we discretized the continuous part into $n=20$ modes with the simple discretization\cite{Craig2005rate} $\omega_k=-\omega_\text{c}\log\frac{k+1/2}{n}$, $g_k=\sqrt{\frac{2\xi\omega_\text{c}}{n\omega_k}}$ ($k=0,\dots,n-1$). The number $n$ was determined by repeating the calculation for increasingly fine discretization until convergence.

\subsection*{FMO}
The supplementary file \texttt{FMO-HS.dat} contains the $8\times 8$ matrix of site energies and couplings\cite{amBusch2011fmo}. The columns of the file \texttt{FMO-specden.dat} contain the values of $\hbar\omega_k$ (in cm$^{-1}$) and $g_k^2$ (dimensionless) for the discrete part of the spectral density.\cite{wendling2000electron}  In the main text, the continuous part of the spectral density was of the form optimized for the B777 complex by Renger and Marcus:\cite{renger2002,Klinger2023}
\begin{equation}\label{eq:B777}
    S(\omega)=\frac{S_0}{s_1+s_2}\sum_{i=1,2} \frac{s_i}{7!\,2\omega_i^4}\omega^3 e^{-(\omega/\omega_i)^{1/2}}
\end{equation}
where $s_1=0.8$, $s_2=0.5$, $\hbar\omega_1=\SI{0.069}{meV}$, and $\hbar\omega_2=\SI{0.24}{meV}$. 
The overall Huang-Rhys factor was set to $S_0=0.42$ for FMO.\cite{adolphs2006fmo} 
The spectral density of bath reorganization energies is releated to $S(\omega)$ by $\Lambda(\omega) = \omega S(\omega)$. We discretized the continuous bath using an equally spaced grid with $n=100$ modes up to $\hbar\omega_\text{max}=\SI{540}{cm^{-1}}$, and checked that the results did not change on including more modes.
The reorganization energy was $\SI{44}{cm^{-1}}$ for the discrete part and $\SI{38}{cm^{-1}}$ for the continuous part of the spectrum.

\begin{figure}[h!]
    \centering
    \includegraphics{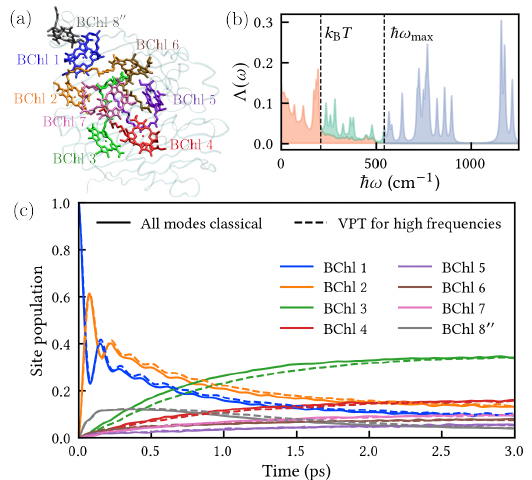}
    \caption{FMO with a more detailed low-frequency part of the spectral density.}
    \label{fig:fmoextra}
\end{figure}

To check that the results are not sensitive to the particular choice of continuous bath, we have also repeated the calculation replacing the continuous part with a spectral density extracted from a modern normal-mode analysis of the full atomistic complex.\cite{klinger2020fmo} We course-grained the Huang-Rhys factors of the 59367 normal modes into 200 effective modes below $\hbar\omega_\text{max}=\SI{540}{cm^{-1}}$. Since Ref.~\onlinecite{klinger2020fmo} reports that the variations of the spectral density between sites are unimportant, we used the site-averaged bath for all sites.
The modified spectral density is shown in Fig.~\ref{fig:fmoextra}(b). Its reorganization energy, $\SI{15}{cm^{-1}}$, is weaker than the bath of the B777 form. For the dynamics [see panel(c)] one only observes a slight difference between the VPT and classical populations for BChl 3, and the conclusions of the main text are unchanged.

\subsection*{LHCII}
The file \texttt{LHCII-HS.dat} contains the couplings from Ref.~\onlinecite{Mueh2010lhcii} with the refined site energies from Ref.~\onlinecite{Mueh2012refined}. To be consistent with previous studies,\cite{Mueh2010lhcii} we included static disorder in the diagonal terms by sampling site energies for each trajectory from the average in \texttt{LHCII-HS.dat} with a standard deviation of $\SI{120}{cm^{-1}}$.

The continuous spectral density was of the same form as in Eq.~\eqref{eq:B777} but with $S_0=0.5$, discretized into $n=200$ modes.\cite{Renger2011lhc2}
The file \texttt{LHCII-specden.dat} contains the discrete spectral density that was constructed as a correction to the continuous form by fitting to the low-frequency flourescence spectrum.\cite{Renger2011lhc2}
The upper cut-off $\hbar\omega_\text{max}$ was recalculated for each instance of the static disorder.

\section*{Supplementary Note 2: Mixed quantum-classical equilibrium}
\begin{figure}[h!]
    \centering
    \includegraphics{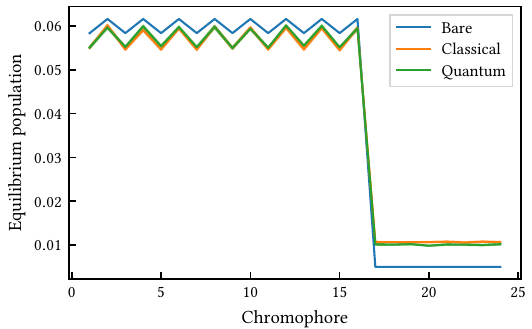}
    \caption{Equilibrium populations of LH2 chromphores computed with classical $(P=1)$ and quantum $(P=8)$ nuclei as described in Supplementary Note 2. Shaded areas denote two standard errors in the mean. Also shown for comparison are the bare populations without any coupling to the vibrational bath.}
    \label{fig:eql}
\end{figure}

To verify that the mixed quantum-classical equilibrium populations in Eq.~\eqref{eq:mixedQC} provide a close approximation to their quantum mechanical counterparts in Eq.~\eqref{eq:Qeq} for LH2, we calculated each type of expectation value explicitly as follows. The quantum mechanical thermal expectation value $\langle |n\rangle\langle n|\rangle$ can be calculated exactly from the path-integral expression
\begin{equation}
    \langle |n\rangle\langle n|\rangle = \lim_{P\to\infty} \frac{\int d^{P}q \, e^{-\frac{\beta}{P}\sum_{i=1}^{P}\frac{P^2}{2\hbar^2\beta^2}(q_i-q_{i+1})^2}\Tr_\text{ex}\left[\prod_{i=1}^{P}e^{-\frac{\beta}{P}V(q_i)} |n\rangle\langle n|\right]}{\int d^{P}q \, e^{-\frac{\beta}{P}\sum_{i=1}^{P}\frac{P^2}{2\hbar^2\beta^2}(q_i-q_{i+1})^2}\Tr_\text{ex}\left[\prod_{i=1}^{P}e^{-\frac{\beta}{P}V(q_i)}\right]}
\end{equation}
where $P$ is the number of imaginary time ring polymer beads.
The corresponding mixed quantum-classical expectation value is obtained by replacing $P=1$ in this expression. We calculated each quantity with Monte Carlo using $10^5$ samples (divided into 10 batches to calculate standard errors of the mean). The quantum expectation values were calculated with $P=8$, which agreed closely with $P=4$. 
The results are shown for each site in Fig.~\ref{fig:eql}. Also shown for comparison are the expectation values for the `bare' excitonic system described by $H_{\rm s}$ without any coupling to the bath. As has been shown previously,\cite{Moix2012} the bath does notably change the equilibrium populations compared to the bare excitonic system. However, the difference between treating the bath with quantum or classical statistics is negligible. For this reason, mixed quantum-classical methods such as MASH that are consistent with Eq.~(5) are bound to agree with an exact quantum mechanical calculation of the chromophore populations at long times (assuming that the mixed quantum-classical dynamics is sufficiently ergodic to reach thermal equilibrium, as it invariably is in these excitonic systems).

\section*{Supplementary Note 3: Wigner distribution}
\begin{figure}
    \centering
    \includegraphics{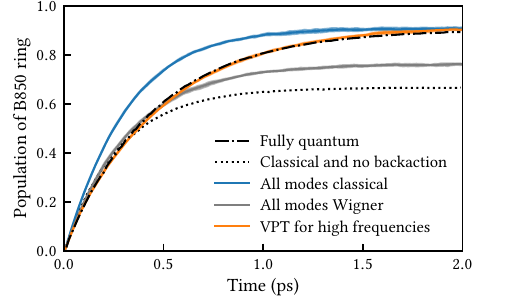}
    \caption{Same as in Figure~\ref{fig:lh2}(c) of the text, but including MASH dynamics starting from a Wigner distribution (gray line). The MASH dynamics starting from a classical distribution reaches the correct mixed quantum-classical equilibrium, which provides a close approximation to the exact quantum mechanical equilibrium for the reasons explained in Supplementary Note 2. However, the long-time dynamics obtained from the Wigner initial conditions reaches an overheated equilibrium, in much the same way as a classical calculation without any back-action (although for different reasons). This problem is solved by treating the high-frequency modes with the VPT and the remainder with classical MASH, which fixes both the long-time equilibrium and the timescale of the relaxation.} 
    \label{fig:wigner}
\end{figure}

Starting from a thermal Wigner distribution effectively initializes each bath mode $k$ at a different elevated temperature corresponding to
\begin{equation}
    \beta_k = \beta \frac{\tanh(\beta\omega_k/2)}{\beta\omega_k/2}.
\end{equation}
The exciton dynamics of LH2 obtained using this initialization is compared to the result obtained starting from a classical Boltzmann distribution (as was done in the main text) in Fig.~\ref{fig:wigner}.

\end{document}